\documentclass[a4paper,11pt]{article}
\usepackage{jheppub} 
\usepackage{lineno}
\usepackage{float}
\usepackage{slashed}
\usepackage{xfrac,bigints}
\usepackage{physics}
\usepackage{subfigure}
\usepackage{tikz}
\usepackage{quantikz}
\usepackage{bm}
\usepackage{hyperref}
\usepackage{url}
\usepackage{color}
\usepackage[normalem]{ulem}

\title{Quantum simulation of real-time current correlators and DIS-inspired observables in the Schwinger model}

\author[a,b]{Kazuki Ikeda,}
\emailAdd{Kazuki.Ikeda@umb.edu}
\affiliation[a]{Department of Physics, University of Massachusetts Boston, Boston, MA 02125, USA}
\affiliation[b]{Center for Nuclear Theory, Department of Physics and Astronomy, Stony Brook University, Stony Brook, New York 11794-3800, USA}

\author[c,d,e]{Zhong-Bo Kang,}
\emailAdd{zkang@physics.ucla.edu}
\affiliation[c]{Department of Physics and Astronomy, University of California,
Los Angeles, CA 90095, USA}
\affiliation[d]{Mani L. Bhaumik Institute for Theoretical Physics, University of California,
Los Angeles, CA 90095, USA}
\affiliation[e]{Center for Quantum Science and Engineering, University of California, Los Angeles, CA 90095, USA}

\author[b,f]{Dmitri E. Kharzeev,}
\emailAdd{dmitri.kharzeev@stonybrook.edu}
\affiliation[f]{Energy and Photon Sciences Directorate, Condensed Matter and Materials Sciences Division,
Brookhaven National Laboratory, Upton, New York 11973-5000, USA}

\author[g,h]{and Wenyang Qian}
\emailAdd{wqian@ccnu.edu.cn}
\affiliation[g]{Institute of Particle Physics and Key Laboratory of Quark and Lepton Physics (MOE),
Central China Normal University, Wuhan, 430079, Hubei, China}
\affiliation[h]{Instituto Galego de F\'isica de Altas Enerx\'ias IGFAE, Universidade de Santiago de Compostela,
E-15782 Galicia-Spain}

\abstract{Hadronic tensors encode the nonperturbative structure of hadrons probed in deep inelastic scattering (DIS), yet their direct evaluation requires real-time evolution that presents a challenge for traditional Euclidean lattice approaches. In this work, we present the first quantum simulation of real-time hadronic current--current correlators in a confining gauge theory, from which DIS-inspired structure functions are extracted as a proof-of-principle demonstration in the Schwinger model, i.e (1+1)-dimensional QED. Using two complementary quantum-simulation strategies -- quantum-circuit and tensor-network methods -- we compute the real-time current–current correlator directly on the lattice and validate our results against exact diagonalization where applicable. From this correlator, we compute the hadronic tensor and determine the longitudinal structure function, the sole nonvanishing DIS observable in two space-time dimensions. Our study demonstrates that quantum simulation offers a viable complementary pathway towards the evaluation of real-time observables  relevant for hadronic structure. It also provides a foundation for extending the  calculations from Schwinger model to other gauge theories.
}

\begin{document}
\maketitle
\flushbottom

\section{Introduction}

Hadronic structure is probed in experiment through the real-time scattering processes. In particular, Deep Inelastic Scattering (DIS) provides one of the most stringent tests of QCD at large momentum transfer and enables the extraction of parton distributions~\cite{RevModPhys.63.615}. From a theoretical perspective, however, DIS cross section remains difficult to compute directly because the relevant observables are defined through \emph{real-time} current–current correlation functions, which lie outside the reach of traditional Euclidean lattice techniques.

Recent progress in universal quantum computing has opened new opportunities in high-energy and nuclear physics~\cite{Bauer:2019qxa, DiMeglio:2023nsa, Fang:2024ple}, as quantum simulations provide access to real-time evolution and open the possibility for calculating dynamical observables that are otherwise inaccessible. In particular, there has been rapid progress on the computation of parton distribution functions~\cite{Lamm:2019uyc, Kreshchuk:2020dla,Qian:2021jxp, LiTianyin:2021kcs,Grieninger:2024cdl,Banuls:2024oxa, Banuls:2025wiq,Kang:2025xpz,Chen:2025zeh,Grieninger:2025mbm}, light-cone distribution amplitudes~\cite{Li:2022lyt,Kang:2025xpz}, fragmentation functions~\cite{LiTianyin:2024nod,Grieninger:2024axp}, hadronic vacuum polarization~\cite{Barata:2024bzk}, and hadronic scattering~\cite{Belyansky:2023rgh,Papaefstathiou:2024zsu,Su:2024uuc,Joshi:2025rha,Schuhmacher:2025ehh,Chai:2025qhf,Davoudi:2025rdv,Farrell:2025nkx,Barata:2025hgx,Barata:2025rjb} using various quantum simulation techniques from quantum circuit to tensor network methods. 
Among them, tensor-network methods provide a powerful nonperturbative tool in $1\!+\!1$ dimensions and have successfully captured the real-time dynamics of both Abelian and non-Abelian gauge theories. They offer controlled access to scattering at moderate energies, but their efficiency depends on the growth of bipartite entanglement. At higher energies or long evolution times, entanglement increases rapidly, restricting existing studies to near-elastic kinematics and preventing reliable access to the genuinely deep-inelastic regime, as demonstrated in recent simulations of meson and baryon scattering in $\mathrm{SU}(2)$ gauge theory~\cite{Barata:2025rjb}.

Traditional lattice QCD has also pursued DIS through Euclidean four-point functions combined with analytic continuation, but extracting real-time information requires solving an inverse Laplace transform, an exponentially ill-conditioned problem. Reconstruction methods such as maximum entropy or Backus–Gilbert inversions yield partial insight at low energies, yet uncertainties proliferate precisely in the high-$Q^2$, small-$x_B$ regime where DIS is most informative~\cite{Liu:1993cv,Liu:1999ak,Liang:2019frk}.

In this work, we present a complementary strategy for computing real-time current correlators and DIS-inspired observables in $1+1$-dimensional confining gauge theories. 
Rather than attempting to simulate physical DIS in QCD directly, we adopt the current--current correlator definition of the hadronic tensor as a mathematical template to define and compute analogous observables in a tractable confining gauge theory, where the full simulation pipeline can be rigorously developed and 
benchmarked.
Instead of relying on full scattering simulations or relying on analytic continuation, we compute the \emph{hadronic tensor itself} directly from its real-time operator definition. We employ two independent nonperturbative approaches: (i) a quantum-simulation algorithm for real-time evolution based on Gauss-law–reduced Hamiltonians, and (ii) tensor-network methods that serve both as a classical benchmark and as a means to explore the regime where entanglement remains tractable. Together, these methods enable the first direct computation of DIS-inspired structure functions in a confining gauge theory from real-time correlation functions. Specifically, we work with the (1+1)-dimensional QED Hamiltonian, the massive Schwinger model~\cite{Schwinger:1962tp,Coleman1976,Susskind:1976jm}. The Schwinger model is particularly suitable as a testbed for quantum simulation since it shares essential features of QCD such as confinement and chiral symmetry breaking, while remaining tractable with tensor networks and near-term quantum algorithms. 

This paper is organized as follows. In Sec.~\ref{sec:DIS}, we introduce the hadronic tensors in the context of DIS. In particular, we present the procedure for obtaining the hadronic tensor and longitudinal structure function. In Sec.~\ref{sec:setup}, we introduce the (1+1)-dimensional QED Hamiltonian and discuss the quantum simulation methods used to calculate the hadronic tensors using quantum circuit and tensor network simulation. In Sec.~\ref{sec:results}, we compare results for the current-current correlator, a key component in the calculation of hadronic tensors, using various simulation strategies. Finally, we evaluate the hadronic tensor and the longitudinal structure function using tensor network with a sufficiently large number of qubits. In Sec.~\ref{sec:conclusion}, we summarize the paper and discuss possible future directions. In Appendix \ref{sec:QC}, we describe quantum circuits for simulating DIS. 

\section{Deep Inelastic Scattering }
\label{sec:DIS}

\begin{figure}[tb]
\centering
\includegraphics[width=0.7\linewidth]{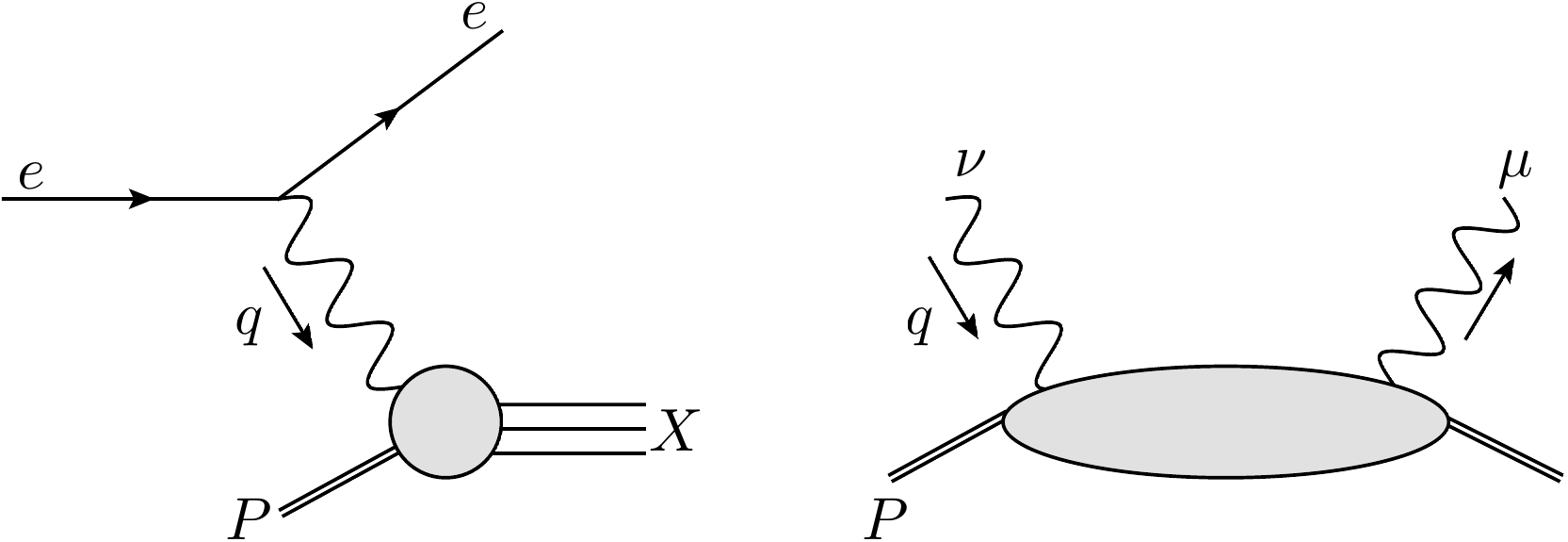}
\caption{Left: Illustration of Deep Inelastic Scattering of an electron and a hadron. The incoming hadron momentum is $P$, and the momentum of the exchanged virtual photon is $q$. Right: The diagram representation of the so-called hadronic tensor, $W^{\mu\nu}(P,q)$.}
\label{f.reaction}
\end{figure}

The deep inelastic scattering (DIS) process
\begin{equation}
    e(\ell) + h(P) \to e(\ell') + X\,,
\end{equation}
illustrated in Fig.~\ref{f.reaction}, plays a central role in revealing the partonic structure of hadrons in strong-interaction physics. A key feature of DIS is that its cross section factorizes into a leptonic tensor $L^{\mu\nu}$ and a hadronic tensor $W^{\mu\nu}$. The leptonic tensor is fully determined by the electromagnetic interaction between the electron and the exchanged virtual photon with momentum $q = \ell - \ell'$, as shown in the left panel of Fig.~\ref{f.reaction}. In contrast, all information about hadron structure and the underlying strong-interaction dynamics resides in the hadronic tensor.

The hadronic tensor is defined through the operator expression~\cite{Collins:2011zzd}
\begin{align}
    W^{\mu\nu}(P,q)
    &= \frac{1}{4\pi} \int d^d y \, e^{iq\cdot y}\,
    \langle h(P) \,|\, 
        J^\mu\!\left(y_1\right)
        J^\nu\!\left(y_0\right)
    \,|\, h(P) \rangle \,,
    \label{e.W}
\end{align}
where $y=y_1 - y_0$. Here $|h(P)\rangle$ denotes the physical hadron state $h$ with momentum $P$, and $J^\mu=\bar{\psi}\gamma^\mu\psi$ is the electromagnetic current.  
The $d$-dimensional space–time coordinate is written inline as $y^\mu \equiv \vec y =  (t, x, \mathbf{y}_t)$, where $\mathbf{y}_t$ is the $(d-2)$-dimensional transverse vector; in $1{+}1$ dimensions, this simply reduces to $ \vec y  = (t, x)$. By translational invariance, the hadronic tensor depends only on the relative separation $y$ between the two current insertions, and not on the arbitrary reference point $y_0$. Consequently, the Fourier transform is taken only with respect to $y$, yielding its dependence on the momentum transfer $q$.

With the constraints of gauge invariance and Lorentz invariance, the hadronic tensor may be decomposed into two structure functions $F_{1,2}(x_B,Q^2)$~\cite{Collins:2011zzd}:
\begin{align}
\begin{aligned}
    W^{\mu\nu}(P,q) =&\,
    \left(-g^{\mu\nu} + \frac{q^\mu q^\nu}{q^2}\right)\!F_1(x_B,Q^2)
     \\[4pt]
    &+
    \left(P^\mu - \frac{P\cdot q}{q^2}\,q^\mu\right)
    \left(P^\nu - \frac{P\cdot q}{q^2}\,q^\nu\right)
    \frac{1}{P\cdot q}\,F_2(x_B,Q^2)\,,
\end{aligned}
\end{align}
where $Q^2 = -q^2$ and the Bjorken variable is $x_B = Q^2/(2P\!\cdot q)$. It is often more illuminating to express the cross section in terms of the transverse and longitudinal structure functions $F_{T,L}$, which are related to $F_{1,2}$ by~\cite{Collins:2011zzd,Kovchegov:2012mbw,Beuf:2017bpd}
\begin{align}
    F_1 = \frac{1}{2x_B}\,F_T\,,
    \qquad
    F_2 = F_T + F_L\,.
\end{align}
Here $F_T$ and $F_L$ correspond to contributions from the transverse and longitudinal polarizations of the virtual photon.  

In $1{+}1$ dimensions the virtual photon has no transverse polarization, so $F_T$ vanishes identically and only $F_L$ survives. 
We stress that this is not merely a technical simplification: in $1{+}1$ dimensions the electromagnetic field carries no propagating dynamical degrees of freedom and the DIS kinematic picture does not apply literally. Indeed, earlier attempts to model DIS-like processes in the Schwinger model required introducing additional external probe fields, precisely because the model does not naturally support the virtual-photon exchange~\cite{Dai:1995ap}. In contrast to those approaches, we compute the current--current correlator directly from its operator definition in Eq.~\eqref{e.W}, using it as a mathematical template to define analogous observables in a lower-dimensional confining gauge theory where quantum simulation methods can be developed and benchmarked. In this spirit, $F_L$ computed here should be understood as a model observable probing the real-time response of the confined spectrum, rather than a phenomenological structure function of QCD.
In this case,
\begin{align}
    F_L(x_B,Q^2)
    = 2x_B (-g_{\mu\nu} W^{\mu\nu})
    = 2x_B \left(W^{11} - W^{00}\right)\,.
    \label{e.FL}
\end{align}

The evaluation of $F_L$ therefore reduces to computing the current–current correlator in coordinate space,
\begin{align}
    \Pi^{\mu\nu}(\vec y,\, \vec y_0)
    \equiv 
    \bra{h(P)} J^\mu(t,x)\,J^\nu(t_0,x_0)\ket{h(P)}\,,
    \qquad t>t_0\,,
    \label{e.Pi}
\end{align}
where $|h(P)\rangle$ is the first excited hadron state with charge zero, i.e.\ the lowest charge–neutral meson.  
Since $F_L$ ultimately depends only on the Lorentz-invariant variables $x_B$ and $Q^2$, the above quantum simulation may be carried out in the hadron’s rest frame. Finally, using Eqs.~\eqref{e.FL} and~\eqref{e.W}, one performs the two-dimensional Fourier transform to obtain the longitudinal structure function $F_L(x_B,Q^2)$.

It is worth noting that the same current--current correlator, but evaluated in the vacuum state $|\Omega\rangle$ rather than a hadronic state, has been computed recently in Ref.~\cite{Barata:2024bzk}:
\begin{align}
    W^{\mu\nu}(q)
    &= \frac{1}{4\pi} \int d^2 y \, e^{iq\cdot y}\,
    \langle \Omega \,|\, 
        J^\mu(y_1)\, J^\nu(y_0)
    \,|\, \Omega \rangle \,.
    \label{e.Wvac}
\end{align}
This object is known as the time-like hadronic vacuum polarization tensor, since $q^2 \equiv Q^2 > 0$. In addition to playing a crucial role in the theoretical determination of the muon's anomalous magnetic moment, it also governs the total cross section for the process $e^+e^- \to X$. Computing the hadronic tensor in both the vacuum and hadronic states is therefore essential for understanding scattering processes in deep-inelastic scattering and in $e^+e^-$ annihilation, respectively.

\section{\label{sec:setup}Quantum simulation for Deep Inelastic Scattering}

Quantum information science (QIS) provides direct access to nonperturbative observables via simulating real-time dynamics in quantum field theory. 
We work within the 1+1 dimensional QED (the Schwinger model)~\cite{Schwinger:1962tp,Coleman1976,Susskind:1976jm} and its Lagrangian density is 
\begin{align}
\label{e.L0}
    \mathcal{L} = -\frac{1}{4}F_{\mu\nu}F^{\mu\nu}+\bar{\psi}(i\gamma^\mu\partial_\mu-g\gamma^\mu A_\mu-m)\psi\,,
\end{align}
where we label the space-time coordinate by $x^\mu=(t,x)$.

On the lattice, the fermion field $\psi(x)$ is discretized and represented using staggered fermion fields~\cite{Borsanyi:2010cj} at each position with the lattice spacing $a$,
\begin{align}\label{e.lattice}
\psi(x=na)=\frac{1}{\sqrt{a}}\left(\begin{array}{cc}
     &\ \chi_{2n}\  \\
     &\ \chi_{2n+1}\
\end{array}\right)\,.
\end{align}
To map the discretized Hamiltonian onto qubits, we apply the \emph{Jordan-Wigner (JW) transformation} to the staggered fermion fields~\cite{Jordan:1928wi}, 
\begin{align}\label{e.jw}
\chi_n=\sigma_{n}^-\prod_{i=1}^{n-1}\left(-i\sigma_{i}^z\right), \quad \chi^\dagger_n=\sigma_{n}^+\prod_{i=1}^{n-1}\left(i\sigma_{i}^z\right) \,,
\end{align}
where $\sigma_i^x, \sigma_i^y, \sigma_i^z$ are the Pauli matrices and $\sigma^{\pm}_i = (\sigma_i^x \pm i \sigma_i^y)/2$ on qubit $i$. We use the following convention for the Dirac matrices: $\gamma^0 = \sigma^z$, $\gamma^1 = i\,\sigma^y$, $\gamma^5=\gamma^0 \gamma^1 = \sigma^x$. The Hamiltonian of the QED model on the lattice of $N$ qubits becomes
\begin{align}
\begin{aligned}
\label{e.H}
H=\frac{1}{4a}\sum_{n=1}^{N-1}\Big[\sigma^x_n \sigma^x_{n+1}+\sigma^y_n \sigma^y_{n+1}\Big]
+\frac{m}{2}\sum_{n=1}^N(-1)^n \sigma^z_n+\frac{a g^2}{2}\sum_{n=1}^{N-1}L^2_n,
\end{aligned}
\end{align}
where $g$ is coupling strength, $a$ is the lattice spacing, and the fermion mass $m$ uses the lattice mass shift convention $m\rightarrow m-ag^2/8$ \cite{Dempsey:2022nys}. With the boundary condition $L_0=0$, the Gauss' law constraint leads to the electric field operator $L_n =  a\sum_{j=1}^n Q_j$ written in terms of the local vector and the bilocal axial charge densities~\cite{Ikeda:2023vfk},
\begin{align}
\label{e.Q}
    Q_n \equiv \,& \bar{\psi}\gamma^0\psi \Rightarrow \frac{\sigma^z_n+(-1)^n}{2a},\\
    Q_{5,n} \equiv \,& \label{e.Q5}\bar{\psi}\gamma^5\gamma^0\psi \Rightarrow \frac{\sigma^x_n\sigma^y_{n+1}-\sigma^y_n\sigma^x_{n+1}}{4a}\,.
\end{align}
For the vector and scalar current-current correlators, they are defined as
\begin{align}\label{e.J}
    J^0_n &= \bar{\psi}_n\gamma^0\psi_n \rightarrow  Q_{2n}+Q_{2n-1} = \frac{1}{2a}\big(\sigma^z_{2n}+\sigma^z_{2n-1}\big)\,, \\
    J^1_n &= \bar{\psi}_n\gamma^1\psi_n \rightarrow 2Q_{5,2n} 
    = \frac{i}{a}\big(\sigma^+_{2n-1}\sigma^-_{2n}-\sigma^-_{2n-1}\sigma^+_{2n}\big)\,,  \\
    J_n &= \bar{\psi}_n\psi_n \equiv \frac{1}{2a}\big(\sigma^z_{2n} - \sigma^z_{2n-1}\big)\,.
\end{align}
It is important to note that the Gauss-law reduction that eliminates all gauge degrees of freedom and produces a purely fermionic long-range Hamiltonian, is only specific to 1+1 dimensions. In higher dimensions, this procedure is not available and one must work with a full gauge-matter system, typically requiring truncation of the gauge field Hilbert space. 

A key ingredient for computing the hadronic tensor in DIS is the matrix element of the current-current correlator evaluated between two spacetime points $\vec y_1=(t_1, x_1)$ and $\vec y_0=(t_0, x_0)$ for a hadron state.
For simplicity, we fix $t_0=0$ and define $\Pi^{\mu\nu}(t, x)$ for the matrix element, where $x=x_1-x_0$ and $t=t_1 -t_0=t_1$. The matrix element can then be written as
\begin{align}\label{e.Pi2}
\begin{aligned}
    \Pi^{\mu\nu}(t, x) &= \langle h|J^\mu(t, x_1) J^\nu(0, x_0)|h \rangle=  \langle h |e^{iHt}J^\mu(0, x_1)e^{-iHt} J^\nu(0, x_0) |h \rangle\\
    &=  e^{iE_ht}\underbrace{\langle h |J^\mu(0, x_1)}_{\langle \psi_L|} | \underbrace{e^{-iHt} J^\nu(0, x_0) |h \rangle}_{|\psi_R\rangle}
    =e^{iE_ht}\langle \psi_L|\psi_R \rangle\,,
\end{aligned}
\end{align}
where $\ket{h}$ is the hadron eigenstate with eigenvalue $E_h$ of the Hamiltonian $H$. 

The current--current matrix elements are thus reduced to overlaps of quantum states after real-time evolution, and both the time evolution and the state overlaps can be directly implemented using quantum simulation. As discussed below Eq.~\eqref{e.W}, in the continuum the hadronic tensor is independent of the reference point $y_0$, or equivalently $x_0$ in Eq.~\eqref{e.Pi2}. On a finite lattice, translational invariance is not exact; nevertheless, for sufficiently large lattices and as long as edge effects are under control, different choices of current insertion points yield equivalent numerical results. In practice, this freedom allows us to place $x_0=0$ at the center of the lattice so that $x_1=x_0+x$, or alternatively to choose symmetric positions $x_0=-x/2$ and $x_1=x/2$ to better utilize the lattice volume. We find that both choices lead to similar numerical results within uncertainties, confirming that finite-size effects are well controlled. For the remainder of the paper, we adopt the simpler choice $x_0=0$, so that $x=x_1$.

After evaluating the matrix elements $\Pi^{\mu\nu}(t, x)$ for all possible values of $x$ and $t$, the hadronic tensor as in Eq.~\eqref{e.W} can be computed through the two-dimensional Fourier transforms in space and time,
\begin{align}
\begin{aligned} \label{e.W2}
W^{\mu\nu}\left(x_B, q^0, q^1\right) &= \int\, d^2y\, e^{iq\cdot y}\, \Pi^{\mu\nu}(t, x) =\int\, dx\, dt\, e^{iq^0t}\,e^{-iq^1x}\, \Pi^{\mu\nu}(t, x)\\
&= \sum_{j=0}^{N_x-1}\, \sum_{k=0}^{N_t-1}\, \Delta t\, \Delta x\, e^{iq^0 k \Delta t}\,e^{-iq^1 j \Delta x}\, \Pi^{\mu\nu}(k \Delta t, j \Delta x).
\end{aligned}
\end{align}
The discrete momenta $q^0$ and $q^1$ are selected from the conjugate grid defined by the simulation parameters, so they lie by construction within the support of the discrete Fourier transform. Specifically, $q^0 = k\,\delta q^0$ and $q^1 = j\,\delta q^1$ with integer $k,j$, where $\delta q^0 = 2\pi/(N_t \Delta t)$ and $\delta q^1 = 2\pi/(N_x \Delta x)$ where $\Delta t$ and $\Delta x$ are the smallest steps in time and position used in the numerical simulation. We restrict to $q^0 < q^1$ to ensure $Q^2 > 0$.
In theory, the continuum limit can be reached by taking $\Delta x\sim a \rightarrow 0$ and $\Delta t \rightarrow 0$. Finally, we compute the longitudinal structure function $F_L$, which is the only non-vanishing structure function in $1{+}1$ dimensions. 
Using the definition of $F_L$ given in Eq.~\eqref{e.FL}, we evaluate it directly from the lattice-computed hadronic tensor as
\begin{align}
    F_L(x_B,Q^2)
    = 2x_B \bigl(W^{11}(x_B,q^0,q^1) - W^{00}(x_B,q^0,q^1)\bigr)\,,
    \label{e.FL2}
\end{align}
where $Q^2 = -q^2 = (q^1)^2 - (q^0)^2$.

\section{\label{sec:results}Hadronic tensor and longitudinal structure function}

In this section, we present numerical results for the hadronic tensor and the longitudinal structure function in the (1+1)-dimensional Schwinger model. The computation involves three main steps: (i) preparing the hadronic state $\ket{h}$, (ii) evaluating the current-current correlators $\Pi^{\mu\nu}(t, x)$, and (iii) obtaining $W^{\mu\nu}$ and $F_L$ via Fourier transforms. These steps provide a first-principles determination of DIS observables using quantum simulation. We first show results for both the quantum circuit (QC) using \texttt{Qiskit}~\cite{qiskit2024} and tensor network (TN) simulations using \texttt{ITensor}~\cite{ITensor:2022} compared with exact diagonalization (ED) for a small amount of qubits ($N=12$). Then we present final results using tensor network up to $N=120$ qubits.

The first major task is the preparation of the hadronic state $\ket{h}$. In this work, we treat the first excited state in the charge-neutral ($Q_{\rm tot}=0$) sector as the hadron. The total charge operator $Q_{\rm tot}$ is defined as $Q_{\rm tot}=\sum_n Q_n$, with the local charge $Q_n$ defined in Eq.~\eqref{e.Q}. 
While identifying the first excited charge-neutral state with a ``hadron'' is a significant simplification relative to QCD, as DIS probes bound states with rich internal partonic structure, this state is the lowest-lying confined meson and the natural choice within the Schwinger model. The current--current correlator evaluated in this state encodes real dynamical information about the confined spectrum, such as how the state responds to a local electromagnetic perturbation as a function of energy and momentum transfer.
On quantum computers, there exist several algorithms to prepare states with fixed quantum numbers. One approach is to work with a Hamiltonian projected onto the $Q_{\rm tot}=0$ sector, which simplifies the problem~\cite{Ikeda:2024rzv}; this strategy is employed in our quantum-circuit simulations. Alternatively, variational methods such as the Hamiltonian variational ansatz~\cite{Wiersema:2020ipa} or variational quantum deflation~\cite{Higgott:2018doo,Chandani:2024xza} have been demonstrated in previous studies~\cite{Qian:2021jxp,LiTianyin:2021kcs,Li:2022lyt,LiTianyin:2024nod}. However, these methods typically require a substantial number of gate operations, making them challenging for current quantum hardware. 

For simulations beyond small system sizes, we rely on tensor-network methods as a scalable quantum information science tool for preparing the hadronic state. Within the tensor-network framework, ground and excited states can be obtained in the matrix product state (MPS) representation using the density matrix renormalization group (DMRG) algorithm~\cite{Schollwock2005,frank-dmrg}, implemented with charge-preserving lattice sites as demonstrated in Refs.~\cite{Barata:2024bzk,Barata:2025hgx,Kang:2025xpz}. The DMRG algorithm is a variational algorithm that optimizes a MPS representation of the ground or low-lying excited states of a one-dimensional quantum system. It proceeds by sweeping through the lattice and locally minimizing the energy within a variational manifold controlled by the bond dimension $\chi$, which sets the amount of entanglement that can be captured. We first obtain the ground state as the vacuum state $\ket{\Omega}$ using DMRG with a sufficient number of around 100 sweeps and a numerical accuracy of $10^{-12}$. The hadronic state $\ket{h}$ is then obtained by projecting out the vacuum MPS and performing a new set of enlarged number of 1000 sweeps until convergence is reached for the first excited state. In both the vacuum and hadron state, we used a maximum bond dimension of $\chi_\mathrm{max}=800$ in the DMRG algorithm, and the resulting states are both well-converged with a MPS bond dimension around $\chi \lesssim 50$ for the vacuum state and $\chi \lesssim 150$ for the hadron state.
\begin{figure}[htbp]
    \centering
    \includegraphics[width=0.85\linewidth]{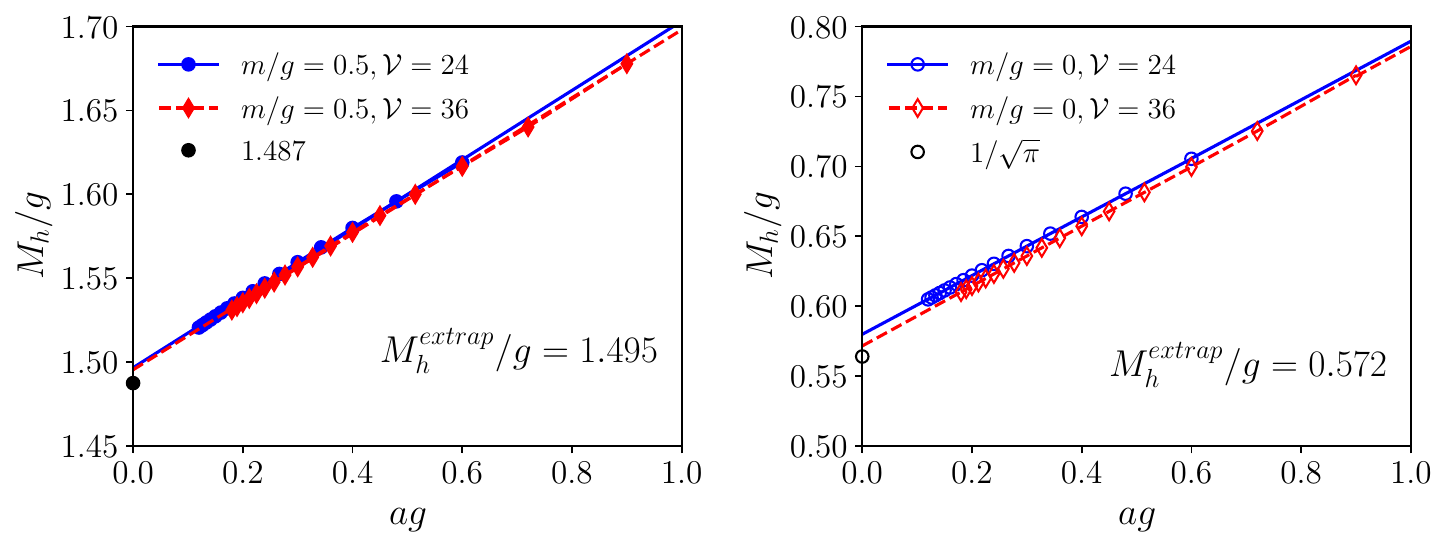}
    \caption{\;Hadron mass evaluated with fixed volume $\mathcal{V}=24$ and 36 with increasing number of qubits up to $N=200$. Extrapolated values at continuum limit $ag=0$ are obtained by a linear fit.}
    \label{f.mass}
\end{figure}

In Fig.~\ref{f.mass}, we present the DMRG results for the hadron mass $M_h = E_h - E_0$, evaluated at fixed physical volumes while decreasing the lattice spacing $a$ (corresponding to increasing the number of qubits $N$). The volumes are fixed at $\mathcal{V} = N a g = 24$ and $36$ respectively, where $E_h = \langle h | H | h \rangle$ and $E_0 = \langle \Omega | H | \Omega \rangle$ denote the energies of the hadronic and ground states, respectively. In principle, the continuum limit is obtained by taking $a \to 0$ and $\mathcal{V} \to \infty$. While performing this limit directly is challenging, reliable results can be obtained by working at sufficiently large volume and extrapolating to vanishing lattice spacing~\cite{Banuls:2013jaa}. In practice, we approach the continuum limit operationally by decreasing the lattice spacing $a \to 0$ at fixed physical volume $V = Nag$, increasing $N$ accordingly, and verifying convergence of physical observables such as the hadron mass to analytically known results. This follows the standard procedure established for the Schwinger model in Ref.~\cite{Banuls:2013jaa}. A full continuum extrapolation for the hadronic-tensor observables is left for future work.

Our results show that the hadron mass is well convergent for sufficiently small $a$, allowing for an extrapolation to $a=0$ to obtain $M_h^{\mathrm{extrap}}$. In the massless limit, we recover the exact analytical result $M_h/g = 1/\sqrt{\pi} \simeq 0.564$ from bosonization~\cite{Schwinger:1962tp}. At finite fermion mass $m/g=0.5$, our results are in good agreement with previous numerical studies based on alternative MPS algorithms~\cite{Banuls:2013jaa}. For the remainder of the paper, we fix the volume to $\mathcal{V}=24$, since for $m/g=0.5$, the results at $\mathcal{V}=24$ and $\mathcal{V}=36$ are virtually indistinguishable, and $\mathcal{V}=24$ is more economical in terms of qubit resources.
At $ag=0.2$, the hadron mass (1.538) already differs from the continuum-extrapolated value (1.495) by only about $3\%$.

To compute the matrix element $\Pi^{\mu\nu}(t,x)$ in Eq.~\eqref{e.Pi2}, we apply local current operators to the hadronic state $\ket{h}$, evolve the state in time $t$, and finally evaluate the overlap. On a quantum computer, these operations are natural and can be implemented using standard real-time evolution with Trotterization together with appropriate measurement protocols. Several strategies have been discussed in Ref.~\cite{Lamm:2019uyc}, including parameterizing the unitary time-evolution operator associated with $\Pi^{\mu\nu}(t,x)$ and extracting the desired matrix element from its derivative. Alternatively, one may employ an ancillary qubit to enable efficient overlap measurements~\cite{Pedernales:2014izf}.
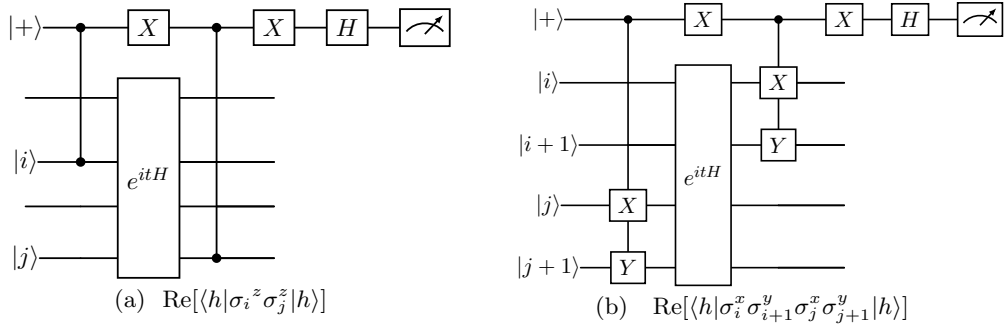
\begin{figure}[htbp]
\centering
\subfigure[\;{$\mathrm{Re}[\langle h| {\sigma_i}^z \sigma_{j}^z|h\rangle]$}]{
\resizebox{0.40\linewidth}{!}{
\begin{quantikz}
\ket{+}&\ctrl{2}&\gate{X}&\ctrl{4}&\gate{X}&\gate{H}&\meter{}\\
&&\gate[4]{e^{itH}}&&\\
\ket{i}&\control{}&\qw&\qw&\qw\\
&&&&\qw\\
\ket{j}&&&\control{}&\qw
\end{quantikz}}
}
\quad
\subfigure[\;
{$\mathrm{Re}[\langle h|\sigma^x_i\sigma^y_{i+1} \sigma^x_j \sigma^y_{j+1}|h\rangle]$}
]{
\resizebox{0.43\linewidth}{!}{\begin{quantikz}
\ket{+}&\ctrl{4}&\gate{X}&\ctrl{2}&\gate{X}&\gate{H}&\meter{}\\
\ket{i}&&\gate[4]{e^{itH}}&\gate{X}&\qw\\
\ket{i+1}&&\qw&\gate{Y}&\qw\\
\ket{j}&\gate{X}&&&\qw\\
\ket{j+1}&\gate{Y}&&&\qw
\end{quantikz}}
}
\caption{\;Quantum circuits for evaluating the components of the two-point correlator $J^\mu(t, x_1) J^\nu(0, x_0)$.}
\label{f.circuit}
\end{figure}

In our study of the $(1+1)$-dimensional Schwinger model, the two-point correlation function $J^\mu(t,x_1)J^\nu(0,x_0)$ defined in Eq.~\eqref{e.J} can be decomposed into a sum of products of simple Pauli operators, which are evaluated individually and combined in postprocessing. The only nontrivial contributions involve terms of the form $\sigma_i^z \sigma_j^z$ and $\sigma_i^x \sigma_{i+1}^y \sigma_j^x \sigma_{j+1}^y$, whose real-part overlaps can be computed using the quantum circuits shown in Fig.~\ref{f.circuit}. Details of the derivation and implementation are provided in App.~\ref{sec:QC}, and the corresponding simulation code is available in Ref.~\cite{code}.

For larger system sizes, we use tensor-network methods. Within the MPS framework, the same sequence of operator application, time evolution, and overlap evaluation can be carried out efficiently. In particular, we employ the time-dependent variational principle (TDVP) algorithm that evolves the MPS by projecting the Schr\"{o}dinger equation onto the tangent space of the MPS manifold at each time step~\cite{Haegeman:2011zz,Haegeman:2015ezw} to perform the real-time evolution. TDVP preserves the norm and energy of the state and avoids the truncation errors that arise in the alternative time-evolving block decimation (TEBD) algorithm that applies a Trotter decomposition of the evolution operator and truncates the bond dimension at each step~\cite{Vidal:2003lvx}.

\begin{figure}[htbp]
    \centering
    \subfigure[\;$\Pi^{\mu\nu}(t, x=0)$]{\includegraphics[width=0.85\linewidth]{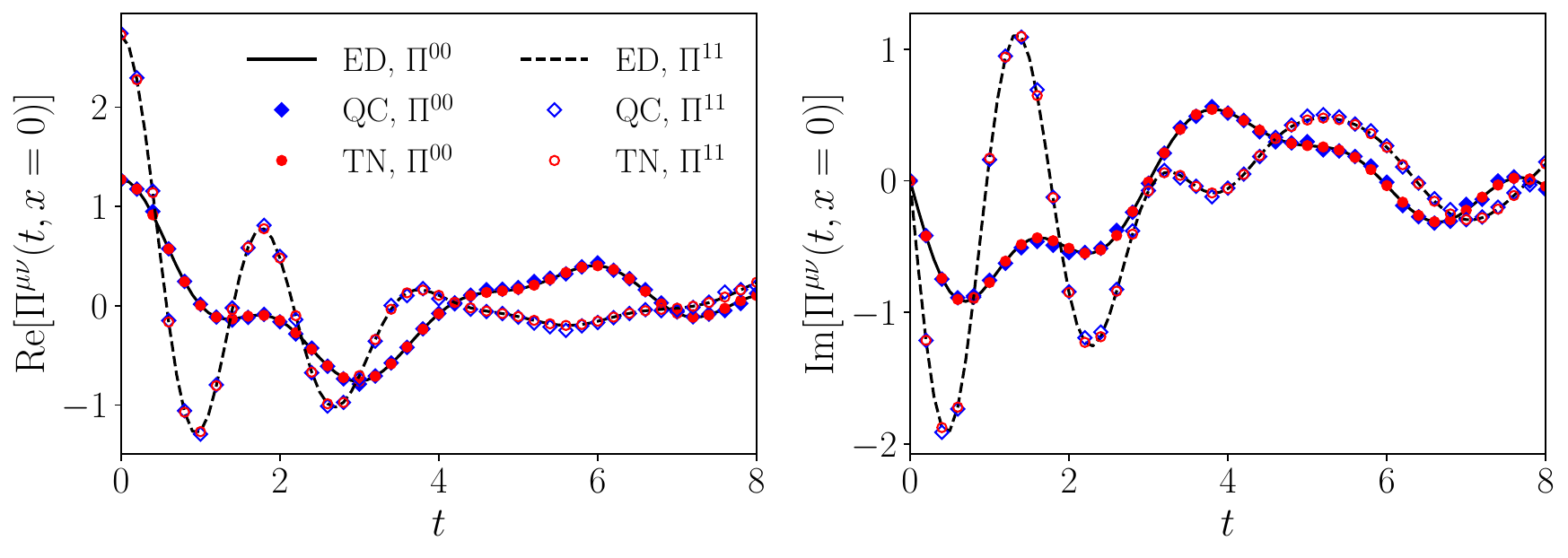}}
    \subfigure[\;$\Pi^{\mu\nu}(t, x=2a)$]{\includegraphics[width=0.85\linewidth]{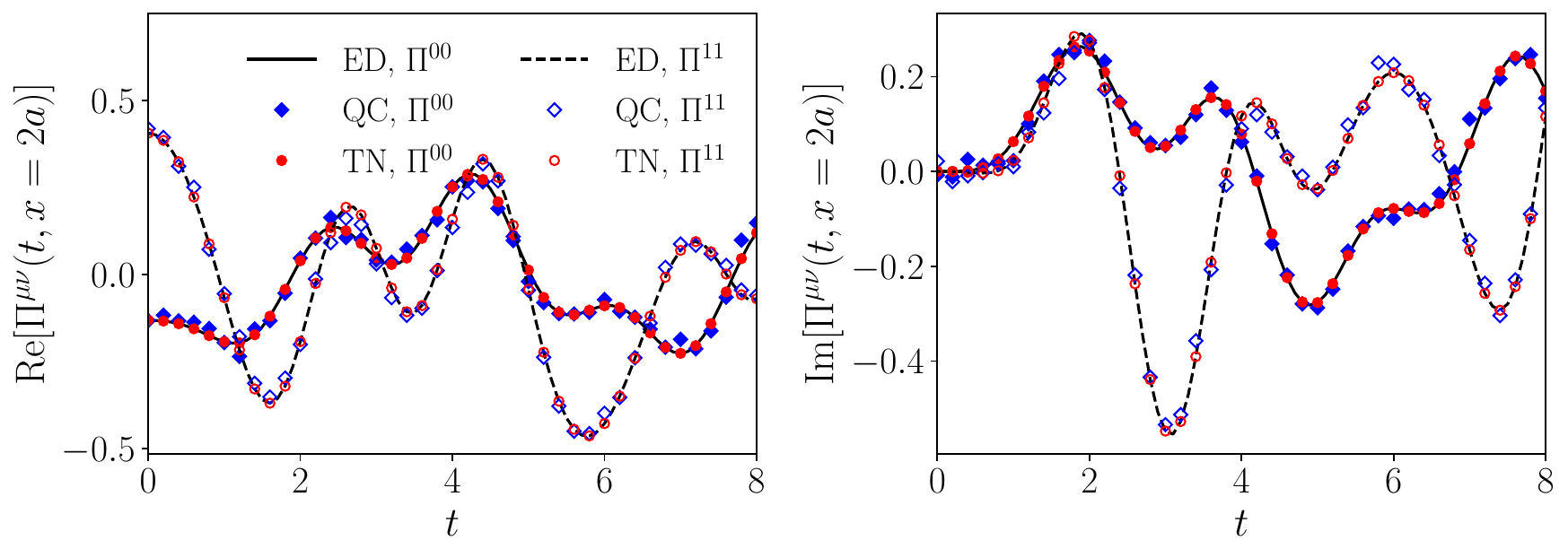}}
    \caption{\;Benchmark comparison of real and imaginary parts of the matrix elements $\Pi^{00}(t, x)$ and $\Pi^{11}(t, x)$ of the hadron state $\ket{h}$ for $N=12$ qubits calculated using exact diagonalization (ED), quantum circuit (QC), and tensor network (TN). Specifically, $m/g=0.5$ and $ag=0.5$ are used. Qubit locations 5 and 6 are used for the center position $x_0=0$.}
    \label{f.Pi_compare}
\end{figure}

Here, we present a benchmark calculation of the matrix elements using both tensor network (TN) and quantum circuit (QC) simulations on a small lattice of $N=12$ qubits, as shown in Fig.~\ref{f.Pi_compare}. For the QC simulation, we implement the time evolution using first-order Trotterization with a time step of $0.1$ and $100,000$ sampling shots on an ideal quantum simulator provided by \texttt{Qiskit}~\cite{qiskit2024}. For the TN simulation, we use the TDVP algorithm provided by \texttt{ITensor}~\cite{ITensor:2022} with a maximal bond dimension of $\chi_\mathrm{max}=250$ and a time step of $0.1$ at a cutoff accuracy of $10^{-12}$. The QC results (shown as square markers) and TN results (shown as circle markers) both agree very well with the true solution obtained with the exact diagonalization results (shown as solid lines) for various correlation separation $x$ and different components of the vector current-current correlators. This agreement validates our quantum simulation approach and provides confidence for extending the calculation to larger lattice sizes. Lastly, we emphasize that the quantum-circuit simulations presented here are performed on an ideal classical simulator using the \texttt{Qiskit statevector} backend. Their purpose is to serve as an independent validation of the tensor-network results at small system sizes ($N=12$), not to claim near-term quantum advantage. The circuits exploit the specific Pauli structure of the 1+1D Schwinger model Hamiltonian and do not directly generalize to higher-dimensional gauge theories without substantially deeper circuits and greater qubit resources. A noise analysis on real quantum hardware, and a resource estimate for higher-dimensional extensions, are important future directions that lie beyond the scope of the present proof-of-principle study.

\begin{figure}[htbp]
    \centering
    \subfigure[\;Hadronic tensors evaluated with $N=80$ and $ag=0.3$. ]{\includegraphics[width=0.85\linewidth]{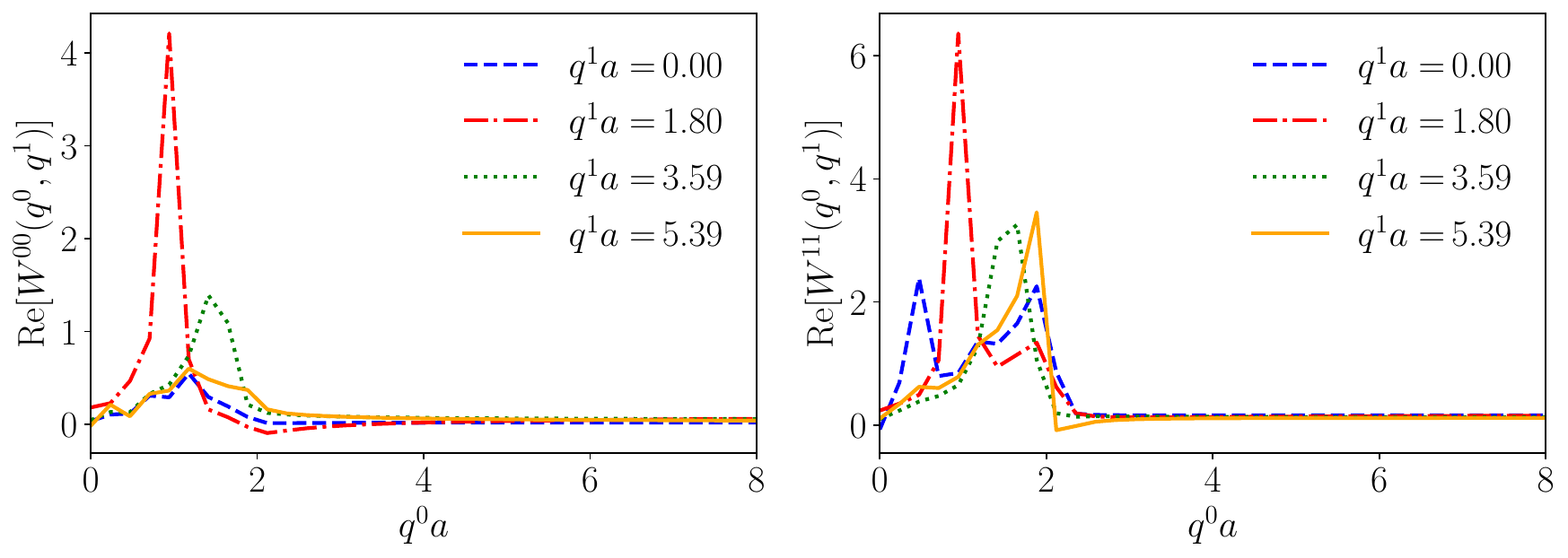}}
    \subfigure[\;Hadronic tensors evaluated with $N=100$ and $ag=0.24$. ]{\includegraphics[width=0.85\linewidth]{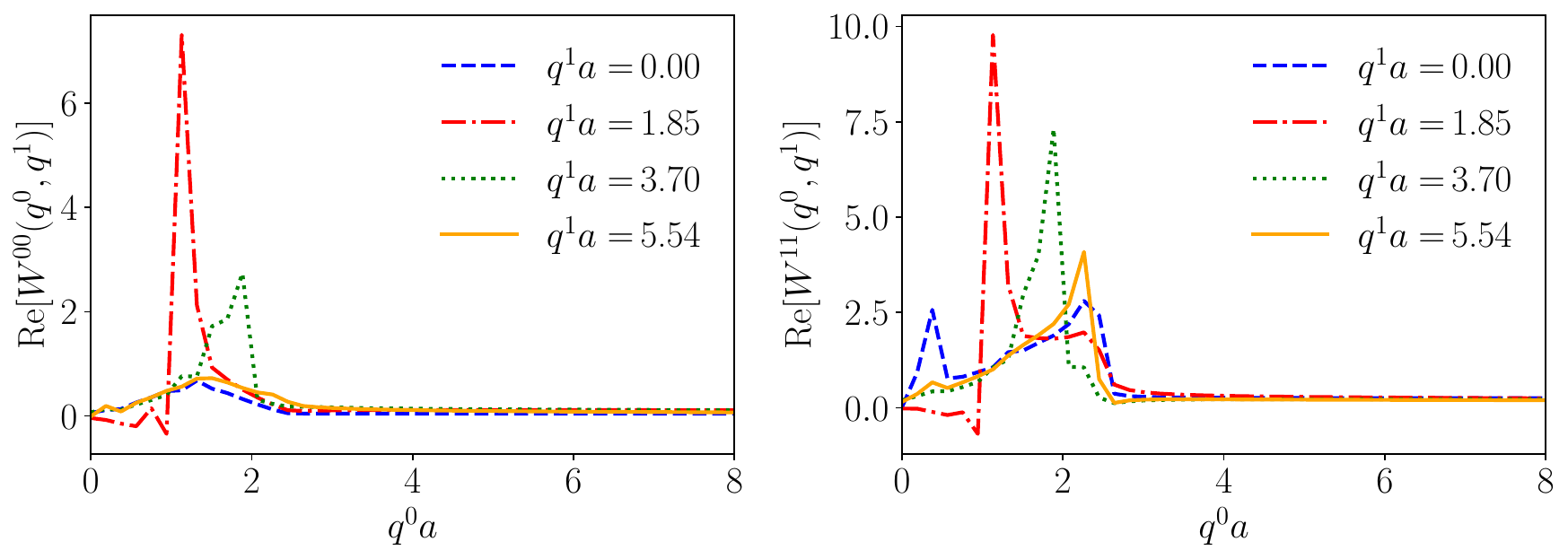}}
    \subfigure[\;Hadronic tensors evaluated with $N=120$ and $ag=0.2$. ]{\includegraphics[width=0.85\linewidth]{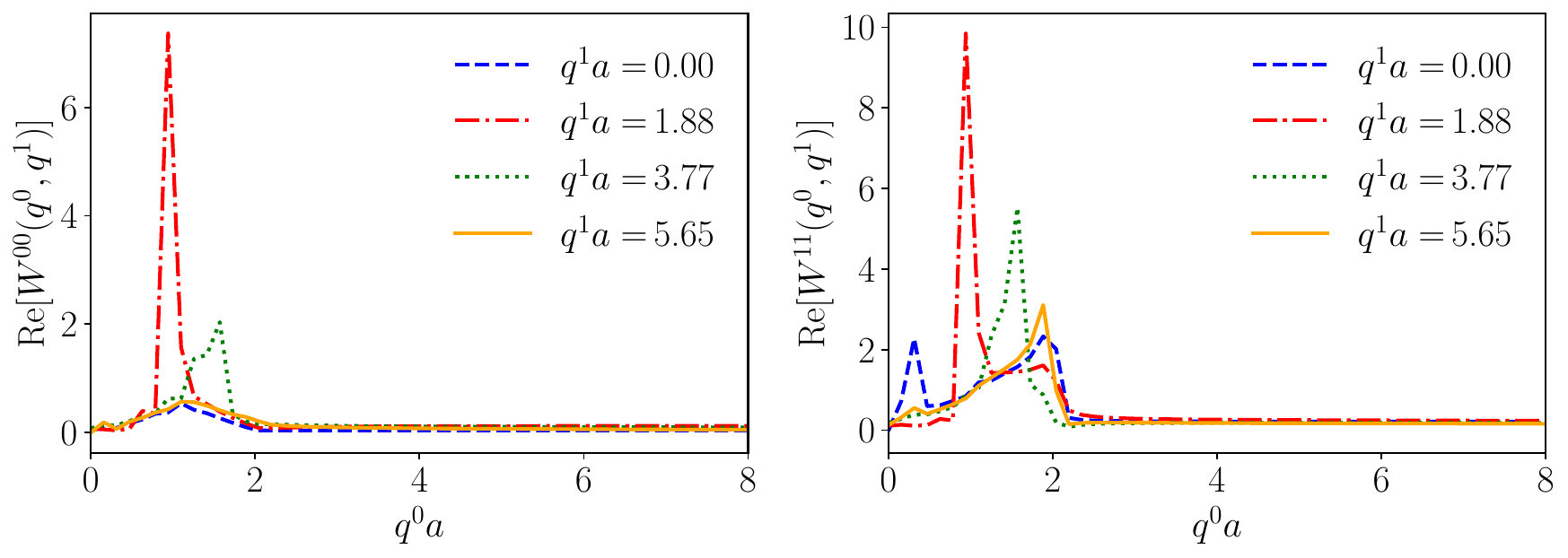}}
    
    \caption{\;Hadronic tensors $W^{00}(q^0,q^1)$ and $W^{11}(q^0,q^1)$ evaluated at fixed longitudinal momentum transfer $q^1$ with decreasing lattice spacing $a$ at constant volume $\mathcal{V}=24$. Here, $m/g=0.5$. }
    \label{f.HT}
\end{figure}

With all the matrix elements available using aforementioned TN methods, we now present our final calculation of the hadronic tensors $W^{00}(x_B,q^0,q^1)$ and $W^{11}(x_B,q^0,q^1)$, as defined in Eq.~\eqref{e.W2}, by performing discrete Fourier transforms in both $q^0$ and $q^1$. In particular, we restrict to the region $q^0 < q^1$ such that $Q^2 > 0$. In Fig.~\ref{f.HT}, we show numerical results at a fixed system size $\mathcal{V}=24$ with increasing number of qubits $N=80,\, 100,\, 120$, corresponding to decreasing lattice spacings $ag=0.3,\, 0.24,\, 0.2$. Despite small variations in the values of $q^1\,a$, controlled by the discretization $\delta q^1=2\pi/(N_x \Delta x)$, both the shapes and peak locations of the hadronic tensors are in good agreement. Notably, in the hadron rest frame, the Bjorken variable $x_B = Q^2 /(2 M_h q^0)$ is heavily controlled by the choices of $q^0$ and $q^1$. To obtain results in the physically relevant region of $0 < x_B < 1$, we use a small lattice spacing down to $ag=0.2$ with $N=120$ qubits. 

\begin{figure}[htbp]
    \centering
    \includegraphics[width=0.42\linewidth]{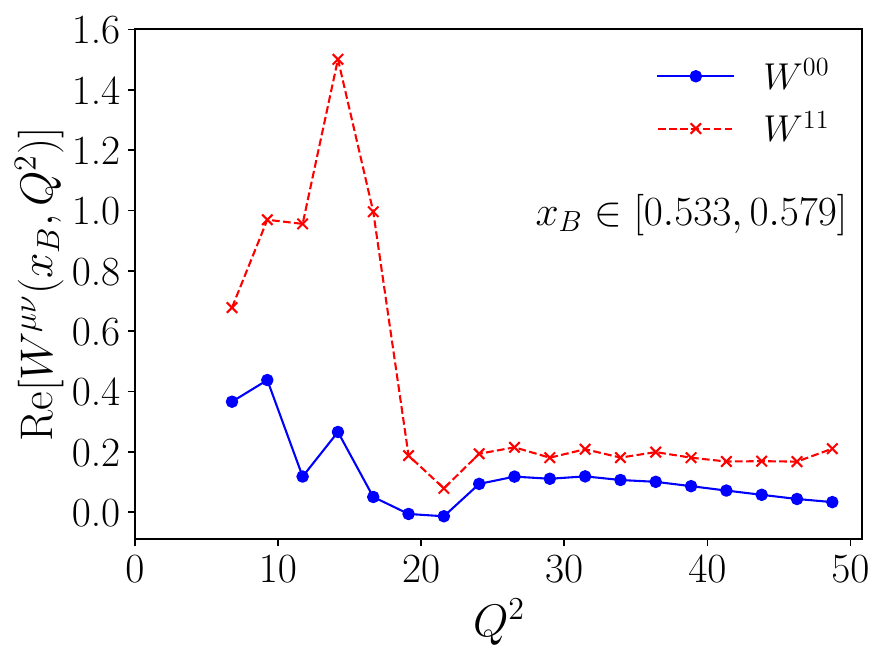}
    \quad\includegraphics[width=0.42\linewidth]{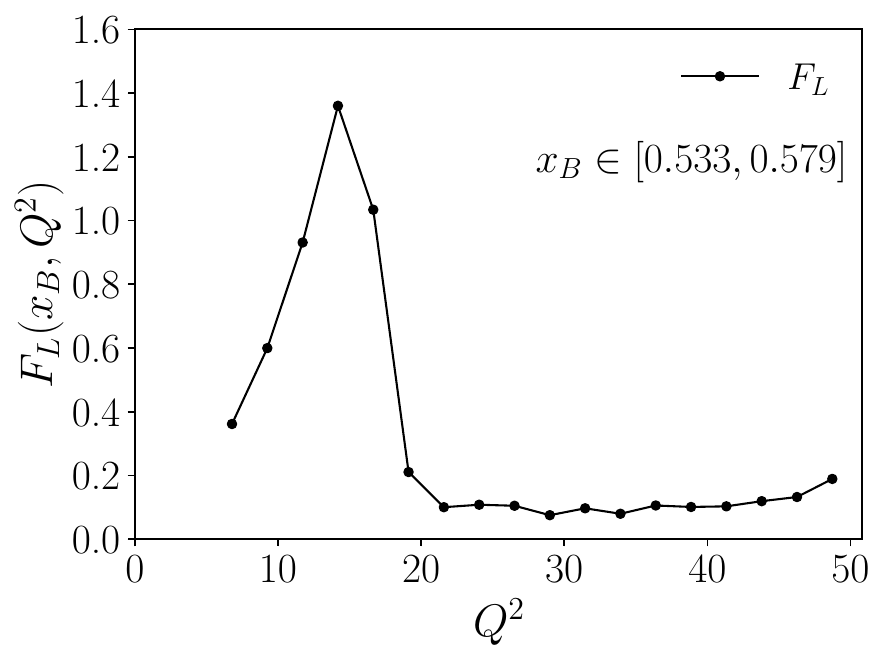}
    \caption{\;Hadronic tensors $W^{\mu\nu}$ and longitudinal structure functions $F_L$ calculated for $N=120$ qubits at fixed volume $\mathcal{V}=24$ using tensor network simulation. Here, $m/g=0.5$ and $ag=0.2$.}
    \label{f.HT_N120}
\end{figure}

Using the smallest lattice spacing of $a=0.2$, we compute in Fig.~\ref{f.HT_N120} both the hadronic tensors and the corresponding longitudinal structure functions according to Eq.~\eqref{e.FL2} for physically relevant values of $0< x_B <1$. In particular, we find the $x_B$ values in the results of the $N=120$ simulation lie approximately in the region around $0.5$, specifically in the range $0.5334 < x_B < 0.579$, primarily due to the specific choices used in our discretizations, $\delta q^0$ and $\delta q^1$. We observe a peak at $Q^2 \sim 15~\mathrm{GeV}^2$ with a magnitude in the range $0.1$--$1$.
Qualitatively, the behavior is consistent with a picture in which longitudinal scattering strength is largest when the virtual probe efficiently couples to the dominant hadron excitations, while at larger momentum transfer the overlap with those excitations diminishes. Notably, Eq.~\eqref{e.FL2} implies that $F_L(x_B, Q^2)$ vanishes as $Q^2 \to 0$. Indeed, since $x_B = Q^2/(2M_h q^0) \to 0$ when $Q^2 \to 0$, we have $F_L(x_B, Q^2) = 2x_B\,(W^{11}-W^{00}) \to 0$ provided the hadronic tensor components $W^{00}$ and $W^{11}$ remain finite, which is the case here. This behavior follows directly from the definition of the longitudinal structure function in the $(1+1)$-dimensional Schwinger model. For reference, in QCD, physically, in the real-photon limit $Q^2 \to 0$ only transverse photon polarizations exist, so the longitudinal cross section $\sigma_L$ vanishes by definition. 
Consequently, $F_L(x_B,Q^2)\propto Q^2\,\sigma_L(x_B,Q^2)$ must also vanish at $Q^2=0$~\cite{Kovchegov:2012mbw,Beuf:2017bpd}.

\begin{figure}[htbp]
    \centering
    \includegraphics[width=0.88\linewidth]{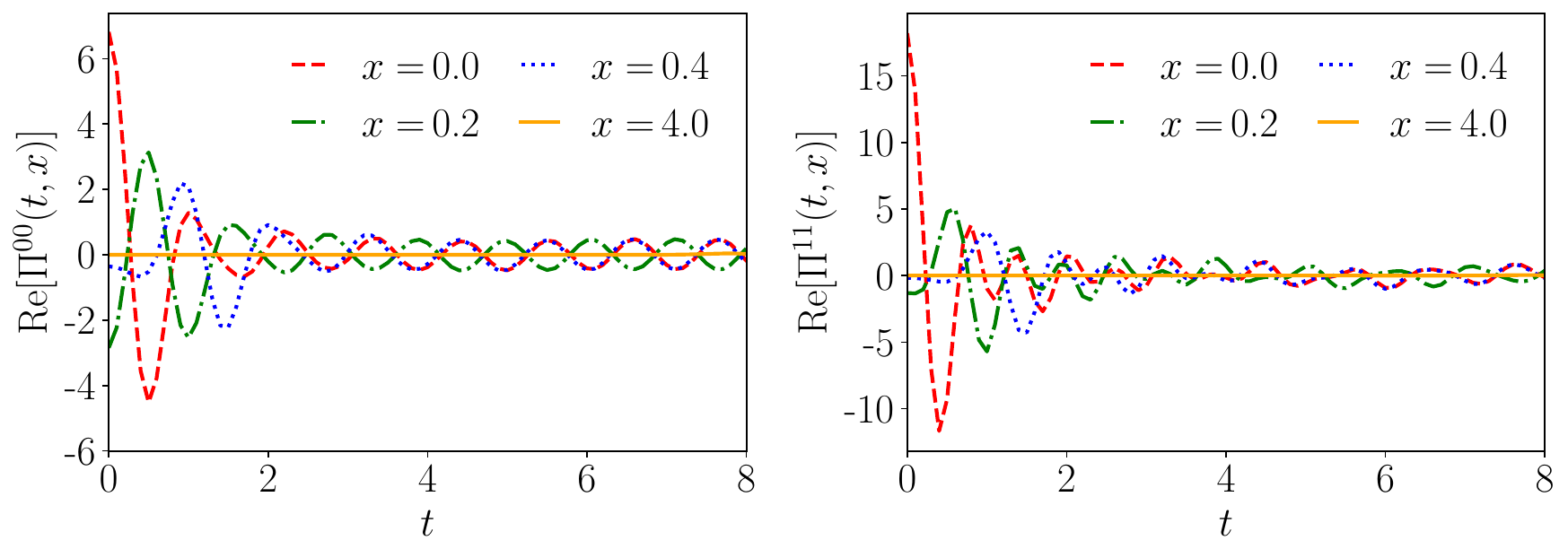}
    \caption{\;Matrix elements $\Pi^{00}(t, x)$ and $\Pi^{11}(t, x)$ evaluated for the hadron state $\ket{h}$ at $N=120$ qubits.}
    \label{f.Pi_N120}
\end{figure}

The main challenge in computing the hadronic tensor and the subsequent $F_L$ is to perform discrete Fourier transforms accurately with many data points covering a wide range of $Q^2$ and with fine lattice resolutions. Ideally, one would like to be able to sample different binned regions of $x_B$ and compute $F_L(x_B, Q^2)$ over various $x_B$'s, which can then be directly used to compare with experimental results. Further investigations are required for smaller lattice spacings and larger system volumes, which are beyond the scope of our current investigation. For the reference, we include the matrix elements $\Pi^{\mu\nu}(t, x)$ used for the calculation of $N=120$ in Fig.~\ref{f.Pi_N120}, where the matrix elements diminish when $x$ is sufficiently large and diverge at $x=0$. The rapid reduction in magnitude with increasing $x$ indicates that the correlator is short-ranged in the hadronic state at the simulated parameters, which is consistent with confinement in the Schwinger model. At $x=0$, the correlator displays a strong enhancement, suggestive of a short-distance contribution.

In the above calculations, the hadron state is always obtained as the first excited state in the charge-neutral sector at fixed momentum. While this provides a clean definition of the hadronic state, one limitation of the fixed-momentum approach on a finite lattice is the restricted kinematic coverage of $x_B$ values. A promising alternative is to construct \emph{wave-packet} states that approximate localized hadrons with a spread in momentum. Such states can be prepared as superpositions of momentum eigenstates within the tensor network framework, or through variational circuits on quantum hardware~\cite{Farrell:2024fit,Farrell:2025nkx, Chen:2025zeh}. The wave-packet approach has the potential to provide smoother access to a wider range of $x_B$ values, reduce finite-volume effects, and bring the lattice construction closer to the physical picture of hadrons probed in scattering experiments. Our preliminary investigation shows similar results using a wave-packet, and we will leave the detailed exploration of this strategy to future work.

\section{\label{sec:conclusion}Conclusion and discussion}

In this work, we have demonstrated a quantum simulation pipeline for computing real-time current--current correlators in the $(1+1)$-dimensional Schwinger model, and shown how DIS-inspired structure functions and hadronic tensors can be extracted from these correlators as a proof-of-principle. Even though the Schwinger model does not realize physical DIS in QCD, the methodology developed here is a valuable and necessary step: it establishes that the full quantum simulation pipeline --- state preparation, real-time evolution, correlator extraction, and Fourier transform to momentum space --- can be reliably benchmarked in a controlled, analytically tractable setting before being extended to more complex gauge theories. By preparing charge-neutral hadronic states using tensor-network techniques and variational quantum algorithms, and by computing real-time current--current correlators through controlled time evolution, we showed how the hadronic tensor and the longitudinal structure function can be obtained from first principles. For small system sizes, results from both tensor-network and quantum-circuit simulations are consistent with exact diagonalization, validating the underlying approach, while tensor-network simulations enable access to larger system sizes.

This study highlights both the feasibility and the promise of extracting hadron structure from quantum simulations. Important future directions include extending these methods to higher-dimensional gauge theories with propagating gauge degrees of freedom and genuine partonic structure, improving state preparation strategies on quantum hardware, employing wave-packet constructions to achieve broader kinematic coverage, and applying these techniques to QCD-relevant observables. By enabling direct access to real-time correlation functions, quantum simulation offers a complementary pathway to hadron structure calculations that are difficult or inaccessible in traditional Euclidean lattice approaches.

\acknowledgments
We are also grateful to Joao Barata, Meijian Li, Tianyin Li, Hongxi Xing, and Ismail Zahed for useful discussions. This work was supported by the U.S. Department of Energy, Office of Science, under Contract No.DE-SC0026415 (KI and DK), by the National Science Foundation under Grant No.~OSI-2328774 (KI)--in particular, on the connection between quantum fundamentals and quantum systems-- and No.~PHY-2515057 (ZK), and by the U.S. Department of Energy, Office of Science, National Quantum Information Science Research Centers, Co-design Center for Quantum Advantage (C2QA) under Contract No.~DE-SC0012704 (DK).
WQ is supported by the European Research Council under project ERC-2018-ADG-835105 YoctoLHC; by the Spanish Research State Agency under project PID2020-119632GB-I00; by Xunta de Galicia (Centro singular de investigacion de Galicia accreditation 2019-2022), by European Union ERDF, and by the Marie Sklodowska-Curie Actions Postdoctoral Fellowships under Grant No. 101109293.

\appendix
\section{\label{sec:QC}Quantum circuits}
In this appendix, we provide the explicit quantum circuits used to evaluate the real-time current--current correlators entering the hadronic tensor
\begin{equation}
\Pi^{\mu\nu}(t, x)=\langle\psi|J^\mu(t, x)J^\nu(0,0)|\psi\rangle \,,
\end{equation}
which is the central observable studied in the main text.
On the lattice, the temporal and spatial components of the current are represented by the charge and axial charge operators,
\begin{align}
\Pi^{00}(t, x=na)=&\langle\psi|Q_{2i}(t)Q_{2j}|\psi\rangle+\langle\psi|Q_{2i}(t)Q_{2j-1}|\psi\rangle\\\notag
&+\langle\psi|Q_{2i-1}(t)Q_{2j}|\psi\rangle+\langle\psi|Q_{2i-1}(t)Q_{2j-1}|\psi\rangle\,,\\
\Pi^{11}(t, x=na)=&4\langle\psi|Q_{5,2i}(t)Q_{5,2j}|\psi\rangle\, ,
\end{align}
where the indices $i,j$ are related to the physical separation $n=i-j$ and $a$ is the lattice spacing.

The purpose of this appendix is to show how these real-time correlators are decomposed into expectation values of Pauli strings and how each term can be measured efficiently using controlled-unitary interferometric circuits.
All circuits shown below are directly implemented in the code provided in Ref.~\cite{code}.

We begin by reviewing a generic interferometric circuit for measuring expectation values of the form
\(\langle\psi|W^\dagger V|\psi\rangle\).
This construction serves as the basic building block for all correlators computed in this work.
In later subsections, the operators \(V\) and \(W\) will be chosen as local Pauli operators or products of Pauli operators appearing in the lattice representations of \(Q_i\) and \(Q_{5,i}\).

\subsection{Basic quantum circuit}
Let $V,W$ be unitary operators and $\ket{\psi}$ be a state. We want to compute $\bra{\psi}W^\dagger V\ket{\psi}$. For this we first see that, by measuring the first qubit of 
\begin{equation}
\label{eq:state}
    \ket{\Psi}=\frac{\ket{0}\otimes V\ket{\psi}+\ket{1}\otimes W\ket{\psi}}{\sqrt{2}}
\end{equation}
in the $X$-basis, we obtain
\begin{align}
    \bra{\Psi}X\otimes I\ket{\Psi}&=\frac{\bra{\psi}W^\dagger V+V^\dagger W\ket{\psi}}{2}=\text{Re}[\bra{\psi}W^\dagger V\ket{\psi}]\,.
\end{align}

This calculation can be computed by the following circuit:
\begin{figure}[H]
    \centering
    \begin{quantikz}
\ket{+}&\ctrl{1}&\gate{X}&\ctrl{1}&\gate{X}&\gate{H}&\meter{}\\
\ket{\psi}&\gate{W}&&\gate{V}\qw&\qw&\qw\\
\end{quantikz}
\end{figure}
Here we used $\ket{+}=\frac{\ket{0}+\ket{1}}{\sqrt{2}}$ and $HZH=X$, where $H$ is an Hadamard gate. 

To compute the imaginary part $\text{Im}[\bra{\psi}W^\dagger V\ket{\psi}]$, we measure the first qubit of $\ket{\Psi}$ in $Y$-basis. This can be seen as follows:
\begin{align}
    \bra{\Psi}Y\otimes I\ket{\Psi}&=\frac{\bra{\psi}iW^\dagger V-iV^\dagger W\ket{\psi}}{2}=i\text{Im}[\bra{\psi}W^\dagger V\ket{\psi}]
\end{align}
Using $SXS^\dagger=Y$, the above computation can be implemented by the following circuit:
\begin{figure}[H]
    \centering
    \begin{quantikz}
\ket{+}&\ctrl{1}&\gate{X}&\ctrl{1}&\gate{X}&\gate{S^\dagger}&\gate{H}&\meter{}\\
\ket{\psi}&\gate{W}&&\gate{V}\qw&\qw&\qw&\qw\\
\end{quantikz}
\end{figure}

\subsection{Time-dependent correlation functions}
To access real-time correlators, the above construction must be extended to include unitary time evolution generated by the lattice Hamiltonian \(H\).
In particular, all components of the hadronic tensor require matrix elements of the form
\(\langle\psi|O(t)O'|\psi\rangle\),
with \(O(t)=e^{itH}Oe^{-itH}\).
The following circuits show how time evolution can be incorporated while preserving the interferometric measurement scheme.

Let $U=e^{-itH}$ be a unitary time evolution, by which the real-time evolution of $V$ is written as $V(t)=e^{itH}Ve^{-itH}$. We are interested in computing $\bra{\psi}WV(t)\ket{\psi}$. For this, we modify the state in Eq.~\eqref{eq:state} as
\begin{equation}
    \ket{\Psi}=\frac{\ket{0}\otimes VU\ket{\psi}+\ket{1}\otimes UW\ket{\psi}}{\sqrt{2}}
\end{equation}
By measuring the first qubit of this state in the $X$-basis, we obtain
\begin{align}
\begin{aligned}
    \bra{\Psi}X\otimes I\ket{\Psi}&=\frac{\bra{\psi}W^\dagger U^\dagger VU+U^\dagger V^\dagger UW\ket{\psi}}{2}\\
    &=\text{Re}[\bra{\psi}W^\dagger V(t)\ket{\psi}]\,,
\end{aligned}
\end{align}
which can be computed by the following circuit:
\begin{figure}[H]
    \centering
    \begin{quantikz}
\ket{+}&\ctrl{1}&\gate{X}&\ctrl{1}&\gate{X}&\gate{H}&\meter{}\\
\ket{\psi}&\gate{W}&\gate{U}&\gate{V}\qw&\qw\\
\end{quantikz}
\end{figure}
When $V$ is a unitary and hermite operator, this circuit can also compute $\text{Re}[\bra{\psi}V(t)W\ket{\psi}]$. For a practical purpose of quantum computation/simulation of physics (e.g. Green's function), this circuit may be enough. 

Suppose we have the following state
\begin{equation}
    \ket{\Psi}=\frac{\ket{0}\otimes V\ket{\psi}+\ket{1}\otimes U^\dagger W^\dagger U\ket{\psi}}{\sqrt{2}}
\end{equation}
and measure its first qubit in $X$-basis. Then we obtain 
\begin{equation}
    \text{Re}[\bra{\psi}U^\dagger W U V\ket{\psi}]=\text{Re}[\bra{\psi}W(t) V\ket{\psi}]\,. 
\end{equation}
The corresponding quantum circuit is given by
\begin{figure}[H]
    \centering
    \begin{quantikz}
\ket{+}&\qw&\ctrl{1}&\gate{X}&\ctrl{1}&\gate{X}&\gate{H}&\meter{}\\
\ket{\psi}&\gate{U}&\gate{W^\dagger}&\gate{U^\dagger}&\gate{V}\qw&\qw\\
\end{quantikz}
\end{figure}
Note that the time evolution $U$ does not act on $\ket{0}\otimes V\ket{\psi}$ since  $U^\dagger U=I$. 

When $V$ is a unitary and hermite operator, this circuit and the previous circuit give the same value in principle. In what follows, we will use the previous circuit since both $V,W$ are Pauli operators, $U$ has a deep gate depth and the fidelity may not be good enough to guarantee $U^\dagger U=I$ on a real quantum computer.  

\subsection{Quantum circuit for $\langle Q_i(t)Q_j\rangle$}
We compute the following two point function:
\begin{equation}
    C(t,x,x_0)=\bra{\psi}O(t,x)O(0,x_0)\ket{\psi}\,,
\end{equation}
where $O(t,x)$ is the time-evolved operator supported at $x$:
\begin{equation}
    O(t,x)=e^{itH}O(0,x)e^{-itH}\,. 
\end{equation}
As a result, \(\Pi^{00}(t, x)\) can be reconstructed entirely from measurements of
\(\langle Z_i(t)Z_j\rangle\),
\(\langle Z_i(t)\rangle\),
and \(\langle Z_j\rangle\).

First we compute the two point operator of charge $O=Q_{n}$. For this, we decompose $C(t,x,x_0)$ as follows:
\begin{align}
\begin{aligned}
    4a^2Q_i(t)Q_j=&Z_i(t)Z_j+Z_i(t)+Z_j\\&+(-1)^iZ_j+(-1)^jZ_i(t)+(-1)^{i+j}\,. 
\end{aligned}
\end{align}
Therefore we can decompose our task into computing the following three operators
\begin{equation}
    \bra{\psi} Z_i(t)Z_j\ket{\psi}\,,~\bra{\psi} Z_i(t)\ket{\psi}\,,~\bra{\psi} Z_j\ket{\psi}\,.  
\end{equation}

Computing $\bra{\psi} Z_i(t)\ket{\psi},~\bra{\psi} Z_j\ket{\psi}$ can be easily done by measuring the states $\ket{\psi(t)}=e^{-itH}\ket{\psi}$ and $\ket{\psi}$ at $j$-site in the $Z$-basis $\ket{0}=\binom{1}{0},~\ket{1}=\binom{0}{1}$. For example, the following circuit compute $\bra{\psi} Z_2(t)\ket{\psi}$ in the four-qubit system
\begin{figure}[H]
    \centering
    \begin{quantikz}
\midstick[4,brackets=none]{$\bra{\psi}Z_2(t)\ket{\psi}$=}&\gate[4]{e^{-itH}}&\qw\\
&\qw&\meter{}\\
&&\\
&&
\end{quantikz}
\end{figure}

In the Schwinger model, however, $\bra{\psi}Z_i(t)\ket{\psi}$ should be constant for any $t$ since the charge is conserved. 

We compute the two point function $\bra{\psi} Z_i(t)Z_j\ket{\psi}$. For example, the following circuit illustrates the way to compute $\text{Re}[\bra{\psi} Z_2(t)Z_4\ket{\psi}]$:
\begin{figure}[H]
    \centering
    \begin{quantikz}
\ket{+}&\ctrl{2}&\gate{X}&\ctrl{4}&\gate{X}&\gate{H}&\meter{}\\
\ket{q_1}&&\gate[4]{e^{+itH}}&\qw&\qw\\
\ket{q_2}&\control{}&\qw&\qw&\qw\\
\ket{q_3}&&&&\qw\\
\ket{q_4}&&&\control{}&\qw
\end{quantikz}
\end{figure}

\subsection{Quantum circuit for $\langle Q_{5,i}(t)Q_{5,j}\rangle$}
Now let us compute the current-current interaction $\langle Q_{5,i}(t)Q_{5,j}\rangle$. This can be done by computing four-point functions 
\begin{align}
\begin{aligned}
    (4a)^2Q_{5,i}Q_{5,j}=&(X_iY_{i+i}-Y_iX_{i+i})(X_jY_{j+i}-Y_jX_{j+i})\\
    =&X_iY_{i+i}X_jY_{j+i}-X_iY_{i+i}Y_jX_{j+i}\\
    &-Y_iX_{i+i}X_jY_{j+i}+Y_iX_{i+i}Y_jX_{j+i}\,.
\end{aligned}
\end{align}
For example, $\text{Re}[\bra{\psi}X_1(t)Y_2(t)X_3Y_4\ket{\psi}]$ can be computed by the following circuit:
\begin{figure}[H]
    \centering
    \begin{quantikz}
\ket{+}&\ctrl{4}&\gate{X}&\ctrl{2}&\gate{X}&\gate{H}&\meter{}\\
\ket{q_1}&&\gate[4]{e^{+itH}}&\gate{X_1}&\qw\\
\ket{q_2}&&\qw&\gate{Y_2}&\qw\\
\ket{q_3}&\gate{X_3}&&&\qw\\
\ket{q_4}&\gate{Y_4}&&&\qw
\end{quantikz}
\end{figure}

In summary, the circuits presented in this appendix provide a complete measurement strategy for reconstructing the hadronic tensor \(\Pi^{\mu\nu}(t, x)\) from real-time quantum simulations.
The temporal component \(\Pi^{00}\) is obtained from two-point \(Z\)-operator correlators, while the spatial component \(\Pi^{11}\) requires four-point correlators involving \(X\) and \(Y\) operators.
By combining the measured real and imaginary parts according to the decompositions given above, all components of the hadronic tensor reported in the main text can be directly obtained from quantum hardware or classical quantum simulators.

\bibliographystyle{JHEP}
\bibliography{ref}

\end{document}